\documentclass[reqno,a4paper,final]{amsart}
\usepackage{amssymb,amstext,amsthm,eucal}
\usepackage{bbold}
\usepackage{array}
\usepackage[latin1]{inputenc}
\usepackage[notref,notcite]{showkeys}
\newcommand{\half}{\tfrac{1}{2}}

\newcommand{\fd}{\mathfrak{d}}
\newcommand{\fg}{\mathfrak{g}}
\newcommand{\fh}{\mathfrak{h}}

\newcommand{\fn}{\mathfrak{n}}

\newcommand{\fso}{\mathfrak{so}}

\newcommand{\SO}{\mathrm{SO}}
\newcommand{\Ort}{\mathrm{O}}
\newcommand{\Cl}{\mathrm{C}\ell}
\newcommand{\Spin}{\mathrm{Spin}}
\newcommand{\Sp}{\mathrm{Sp}}

\newcommand{\SL}{\mathrm{SL}}
\newcommand{\SU}{\mathrm{SU}}
\newcommand{\U}{\mathrm{U}}
\newcommand{\EE}{\mathbb{E}}
\newcommand{\RR}{\mathbb{R}}
\newcommand{\CC}{\mathbb{C}}
\newcommand{\HH}{\mathbb{H}}
\newcommand{\tS}{\mathbb{S}}
\newcommand{\ZZ}{\mathbb{Z}}

\newcommand{\eH}{\mathcal{H}}
\newcommand{\eL}{\mathcal{L}}

\newcommand{\eV}{\mathcal{V}}

\DeclareMathOperator{\id}{id}
\DeclareMathOperator{\AdS}{AdS}
\DeclareMathOperator{\NW}{NW}

\newcommand{\be}{\boldsymbol{e}}
\newcommand{\bv}{\boldsymbol{v}}

\newcommand{\bz}{\boldsymbol{z}}
\newcommand{\bzero}{\boldsymbol{0}}

\newcommand{\bH}{\boldsymbol{H}}

\newcommand{\1}{\mathbb{1}}

\newcommand{\MUNCH}[1]{\relax}
\allowdisplaybreaks[1]
\begin{document}
\title[Supergravity vacua and lorentzian Lie groups]{Supergravity
  vacua and lorentzian Lie groups}
\author[Chamseddine]{Ali Chamseddine}
\author[Figueroa-O'Farrill]{José Figueroa-O'Farrill}
\author[Sabra]{Wafic Sabra}
\address[AC,WS]{
  \begin{flushright}
    Center for Advanced Mathematical Sciences\\
    American University of Beirut, Lebanon
  \end{flushright}
}
\email{chams@aub.edu.lb,ws00@aub.edu.lb}
\address[JF]{
  \begin{flushright}
    School of Mathematics\\
    University of Edinburgh, Scotland
  \end{flushright}
}
\email{j.m.figueroa@ed.ac.uk}
\thanks{CAMS/03-05, EMPG-03-11}
\begin{abstract}
  We classify maximally supersymmetric backgrounds (vacua) of chiral
  (1,0) and (2,0) supergravities in six dimensions and, by reduction,
  also those of the minimal $N{=}2$ supergravity in five dimensions.
  Up to R-symmetry, the (2,0) vacua are in one-to-one correspondence
  with (1,0) vacua, and these in turn are locally isometric to Lie
  groups admitting a bi-invariant lorentzian metric with anti-selfdual
  parallelising torsion, which we classify.  We then show that the
  five-dimensional vacua are homogeneous spaces arising canonically as
  the spaces of right cosets of spacelike one-parameter subgroups.
\end{abstract}
\maketitle

\tableofcontents

\section{Introduction}

This paper is concerned with maximally supersymmetric backgrounds of
minimal supergravity theories in dimensions five and six; that is,
chiral six-dimensional supergravity theories of the type $(1,0)$ and
$(2,0)$ and the minimal $N{=}2$ five-dimensional supergravity
theories.  Much progress has been made recently in dimension five
\cite{GGHPR} in the determination of supersymmetric backgrounds by
rewriting the conditions for the existence of a Killing spinor in
terms of bispinors of the Killing spinor, that is, the differential
forms in the Fierz decomposition of the square of a Killing spinor.
Together with a knowledge of the orbit structure of the spinor
representation, in particular the isomorphism type of the stabiliser,
this method has proven useful in characterising the supersymmetric
solutions in terms of bundle constructions and other standard
geometric constructions.  Nevertheless and despite vigorous effort
\cite[Section 5]{GGHPR} there does not exist a proven list of the
maximally supersymmetric vacua of the minimal $N{=}2$ supergravity.
The list in \cite[Section 5.4]{GGHPR} includes flat space, a symmetric
plane wave discovered by Meessen \cite{Meessen}, the hitherto unknown
Gödel solution, and a one-parameter family of solutions interpolating
between $\AdS_2 \times S^3$ and $\AdS_3 \times S^2$ and which can be
interpreted \cite{GMT} as near-horizon geometries of supersymmetric
rotating black holes \cite{KRW}.  In addition to these
five-dimensional vacua, there are other conjecturally maximally
supersymmetric vacua mentioned in \cite[Section 5.4]{GGHPR} which have
yet to be identified.  One of the aims in the present paper is to
clarify this situation.

The vacua in the above list have been studied from a variety of points
of view \cite{LMO8,ALO} and in particular they have been shown to be
related by dimensional reduction from the minimal chiral supergravity
in six dimensions \cite{LMO8}.\footnote{The Gödel solution was not
  discussed in \cite{LMO8}, but the authors of that paper subsequently
  showed that it can be obtained by reducing the six-dimensional
  symmetric plane wave of \cite{Meessen} (private communication).} The
existence of so many Kaluza--Klein reductions preserving all the
supersymmetry is a very unusual phenomenon deserving of a conceptual
explanation.  In this paper we will provide such an explanation and as
a result we will be able to give a list of all possible
five-dimensional vacua.  Our results will be phrased purely in terms
of Lie algebraic data.

The paper is organised as follows.  In Section~\ref{sec:sugr} we
describe the chiral six-dimensional supergravities of types $(1,0)$
and $(2,0)$.  In Section~\ref{sec:vacua6} we show that the $(1,0)$
vacua (up to local isometry) are in one-to-one correspondence with
six-dimensional Lie groups admitting a lorentzian metric and
anti-selfdual parallelising torsion.  Moreover we show that up to the
action of the R-symmetry group, the $(2,0)$ vacua are in one-to-one
correspondence with the $(1,0)$ vacua.  In Section~\ref{sec:lie} we
determine the six-dimensional anti-selfdual lorentzian Lie groups.
The corresponding vacua are given by flat space, $\AdS_3 \times S^3$
and Meessen's symmetric plane wave.  In Section~\ref{sec:vacua5} we
discuss the Kaluza--Klein reduction to five dimensions of these vacua.
We observe that because the six-dimensional vacua are parallelised Lie
groups the space of right cosets of \emph{any} (spacelike)
one-parameter subgroup is a smooth maximally supersymmetric vacuum
solution of the \emph{minimal} $N{=}2$ supergravity theory in five
dimensions.  We therefore classify such subgroups and hence such
vacua.  Indeed we find other reductions in addition to the list in
\cite[Section 5.4]{GGHPR}; although it still remains to identify them
with the extra possibilities in \cite{GGHPR}.

For a discussion of lorentzian Lie groups in the construction of type
II supergravity backgrounds see \cite{JMFPara,KYPara}.

\section*{Note added}

While this paper rested undisturbed in our computer hard drives, the
paper \cite{GMR} appeared where all supersymmetric backgrounds of the
(1,0) supergravity theory are determined and the maximally
supersymmetric ones classified.  This last result is obtained using
the methods of \cite{FOPMax,FOPPluecker}, whereas in the present paper
we employ a Lie theoretic method which allows in addition to obtain
all the five-dimensional vacua by reduction.

\section{$(1,0)$ and $(2,0)$ supergravities in six dimensions}
\label{sec:sugr}

We describe the field content and Killing spinor equations of $(1,0)$
\cite{NishinoSezgin10} and $(2,0)$ \cite{Townsend20,Tanii20} chiral
supergravities in six dimensions.  We start as usual by describing the
relevant spinorial representations, which in signature $(1,5)$
correspond to symplectic Majorana--Weyl spinors.  More precisely, the
spin group $\Spin(1,5)$ is isomorphic to $\SL(2,\HH)$, whence the
irreducible spinorial representations are quaternionic (i.e.,
pseudoreal) of complex dimension $4$.  There are two inequivalent
representations $S_\pm$ which are distinguished by their chirality.
Let $S_1$ denote the fundamental representation of $\Sp(1)$: it is a
quaternionic representation of complex dimension $2$, and similarly
let $S_2$ denote the fundamental representation of $\Sp(2)$, which is
a quaternionic representation of complex dimension $4$.  The tensor
products $S_+ \otimes S_1$ and $S_+ \otimes S_2$ are complex
representations of $\Spin(1,5) \times \Sp(1)$ and $\Spin(1,5) \times
\Sp(2)$, respectively, with a real structure.  We will let
\begin{equation*}
  S = [S_+ \otimes S_1] \qquad\text{and}\qquad \tS = [S_+ \otimes
  S_2]
\end{equation*}
denote the underlying real representations.  Clearly $S$ is a real
representation of dimension $8$ and $\tS$ is a real representation of
dimension $16$.  The reality condition corresponds to the symplectic
Majorana condition.

The field content of minimal $(1,0)$ supergravity in six dimensions
consists of a metric $g$, an anti-selfdual three-form $H$ and a
gravitino which is a one-form with values in the spinor bundle
associated to $S$, which we will also denote $S$.  As a check, notice
that there is a match of physical degrees of freedom, there being 12
bosonic and 12 fermionic.

On the other hand, the field content of minimal $(2,0)$ supergravity
consists of a metric $g$, a $V$-valued anti-selfdual three-form
$\bH$, where $V$ is the five-dimensional real representation of
the R-symmetry group $\Sp(2)\cong \Spin(5)$, and a gravitino which is
a one-form with values in $\tS$, the spinor bundle associated to
$\tS$.  Again we check that there are 24 physical bosonic and 24
physical fermionic degrees of freedom.

Next we discuss the Killing spinor equations.  In $(1,0)$ supergravity
let $\varepsilon$ be a section of $S$, and the Killing spinor equation
is
\begin{equation}
  \label{eq:conn10}
  D_\mu \varepsilon = \nabla_\mu \varepsilon + \tfrac18 H_\mu^{ab}
  \Gamma_{ab} \varepsilon = 0~,
\end{equation}
where $\nabla$ is the spin connection.  Notice that $D$ is in fact a
spin connection with torsion three-form $H$.

In $(2,0)$ supergravity, let $\varepsilon$ be a section of $\tS$.  The
Killing spinor equation is
\begin{equation}
  \label{eq:conn20}
  D_\mu \varepsilon = \nabla_\mu \varepsilon + \tfrac18 H_\mu^i{}^{ab}
  \Gamma_{ab} \gamma_i \varepsilon = 0~,
\end{equation}
where we have chosen an orthonormal basis $\be_i$ for $V$, so that
$\bH = H^i \be_i$ and $\gamma_i$ are the corresponding generators of
$\Cl(V)$.

The equations of motion consist of the Einstein equations relating the
Ricci tensor of $g$ to the energy-momentum tensor of the three-forms,
and the fact that these three-forms are closed.  Notice that in
$(2,0)$ supergravity, the anti-selfduality of the $H^i$ imply that
$H^i \wedge H^j = 0$ for all $i,j$.

Maximal supersymmetry implies that the connections $D$ acting on $S$
or $\tS$ should be flat.  In the case of $(1,0)$ supergravity, $D$ is
a spin connection with torsion and maximally supersymmetric solutions
correspond to six-dimensional lorentzian manifolds admitting a flat
metric connection with anti-selfdual closed torsion three-form.  We
will see that this means that the manifold is locally isometric to a
Lie group with a bi-invariant lorentzian metric.\footnote{The
  observation that Lie groups with bi-invariant lorentzian metrics
  yield $(1,0)$ vacua was made independently by Meessen and Ortín
  (private communication).}  In the case of $(2,0)$ supergravity, $D$
does not have such an obvious geometrical interpretation, but we will
see below that, up to the natural action of the R-symmetry group on
maximally supersymmetric solutions, the $(2,0)$ vacua are in
one-to-one correspondence with the $(1,0)$ vacua.  In more concrete
terms, we will show that a $(2,0)$ vacuum can be R-transformed to one
where at most one $H^i$ is nonzero.  The flatness equations then
reduce to those in $(1,0)$ supergravity.

\section{Maximally supersymmetric solutions}
\label{sec:vacua6}

In this section we will study the flatness of the spinor connections
\eqref{eq:conn10} and \eqref{eq:conn20}.

\subsection{Vacua of $(1,0)$ supergravity}
\label{sec:10}

The curvature of the $(1,0)$ connection \eqref{eq:conn10} is given by
\begin{equation*}
  \begin{aligned}[m]
   - [D_\mu, D_\nu] &= - [\nabla_\mu + \tfrac18 H_{\mu\,ab}\Gamma^{ab},
    \nabla_\nu + \tfrac18 H_{\nu cd}\Gamma^{cd}]\\
    &= \tfrac14 \left( R_{\mu\nu\, ab} - \tfrac14 \nabla_{[\mu}
      H_{\nu]\,ab} - \tfrac18 \left(H_{\mu\,ac} H_\nu{}^c{}_b -
        H_{\nu\,ac} H_\mu{}^c{}_b \right) \right) \Gamma^{ab}~,
  \end{aligned}
\end{equation*}
where we have used that $\Gamma^a \Gamma^b + \Gamma^b \Gamma^a = 2
g^{ab}\1$.  The quantity
\begin{equation}
  \label{eq:curv10}
  R_{\mu\nu\, ab} - \tfrac14 \nabla_{[\mu} H_{\nu]\,ab} - \tfrac18
  \left(H_{\mu\,ac} H_\nu{}^c{}_b - H_{\nu\,ac} H_\mu{}^c{}_b \right)
\end{equation}
is the curvature of a metric connection with torsion three-form $H$,
whose flatness implies that the spacetime is locally isometric to a
Lie group admitting a bi-invariant metric with parallelising torsion
$H$.  This is a well-known result due to Wolf \cite{Wolf1,Wolf2} based
on earlier work of Élie Cartan and Schouten
\cite{CartanSchouten1,CartanSchouten2}.

Let us sketch how this arises.  The vanishing of the curvature
\eqref{eq:curv10} actually implies several independent equations
corresponding to the decomposition of the curvature tensor into
different algebraic types.  We first rewrite the flatness condition
without reference to the local frame:
\begin{equation*}
  R_{\mu\nu\rho\sigma} - \tfrac14 \nabla_{[\mu} H_{\nu]\rho\sigma}
  - \tfrac18 \left(H_{\mu\rho\tau} H_\nu{}^\tau{}_\sigma -
    H_{\nu\rho\tau} H_\mu{}^\tau{}_\sigma \right) = 0~.
\end{equation*}
The crucial observation is that whereas $R_{\mu\nu\rho\sigma}$
satisfies the algebraic Bianchi identity $R_{[\mu\nu\rho]\sigma}=0$
the other terms do not in general.  Therefore in the first instance,
the above equation breaks into two equations: one sets
$R_{\mu\nu\rho\sigma}$ equal to the component of
\begin{equation*}
  T_{\mu\nu\rho\sigma} := \tfrac14 \nabla_{[\mu} H_{\nu]\rho\sigma}
  + \tfrac18 \left(H_{\mu\rho\tau} H_\nu{}^\tau{}_\sigma -
    H_{\nu\rho\tau} H_\mu{}^\tau{}_\sigma \right)
\end{equation*}
which obeys the algebraic Bianchi identity, and the other equation
says that
\begin{equation*}
  T_{[\mu\nu\rho]\sigma}=0~.
\end{equation*}
Decomposing this last equation further into algebraic types, and using
the fact that $dH=0$ we obtain two equations: the first says that
$\nabla H = 0$ and the second is the Jacobi identity for $H$:
\begin{align}
  H_{[\mu\rho}{}^\tau H_{\nu]\tau\sigma}  = 0~.
\end{align}
Indeed, since $H$ is parallel, so is the Riemann tensor $R$, whence
the spacetime is locally symmetric.  This means that we can work at a
point in the spacetime, on whose tangent space the metric $g$ induces
a lorentzian scalar product and $H$ induces a Lie bracket compatible
with the metric.  In other words, we have the structure of a Lie
algebra with an invariant metric on the tangent space at any point.
This implies that the spacetime is locally isometric to a Lie group
admitting a bi-invariant metric and whose parallelising three-form $H$
is anti-selfdual; indeed, the left-invariant vector fields are
precisely those vector fields which are covariantly constant with
respect to the connection $D$.

Finally, the Riemann curvature is further given in terms of $H$ by
\begin{equation}
  \label{eq:flat10}
  R_{\mu\nu\rho\sigma} = \tfrac18 \left(H_{\mu\rho\tau}
    H_\nu{}^\tau{}_\sigma - H_{\nu\rho\tau} H_\mu{}^\tau{}_\sigma
  \right)~.
\end{equation}

\subsection{Vacua of $(2,0)$ supergravity}
\label{sec:20}

Similarly, the curvature of the $(2,0)$ connection \eqref{eq:conn20}
is
\begin{equation*}
  \begin{aligned}[m]
    - [D_\mu, D_\nu] &= - [\nabla_\mu + \tfrac18 H^i_{\mu\,ab}\Gamma^{ab}
    \gamma_i,\nabla_\nu + \tfrac18 H^j_{\nu cd}\Gamma^{cd}\gamma_j]\\
    &= \tfrac14 R_{\mu\nu}{}^{ab} \Gamma_{ab} - \tfrac1{16}
    \nabla_{[\mu} H^i_{\nu]\,ab} \Gamma^{ab} \gamma_i - \tfrac1{64}
    H^i_{\mu\,ab} H^j_{\nu cd} [\Gamma^{ab}\gamma_i.
    \Gamma^{cd}\gamma_j]
  \end{aligned}
\end{equation*}
Flatness implies the vanishing of all the independent components of
the curvature.  Using that $\gamma_i \gamma_j + \gamma_j \gamma_i = 2
\delta_{ij}\1$, we obtain the following set of equations, which are
equivalent to the flatness of the connection $D$:
\begin{align}
  R_{\mu\nu\,ab} - \tfrac18 \left( H^i_{\mu\, ac} H^i_{\nu\, db}
  - H^i_{\nu\, ac} H^i_{\mu\, db}\right) g^{cd} &=0
  \label{eq:riemann}\\
  \nabla_\mu H^i_{\nu\, ab} - \nabla_\nu H^i_{\mu\, ab} &= 0
  \label{eq:dH}\\
  H^{[i}_{\mu\, ab} H^{j]}_{\nu\, cd} \epsilon^{abcdef} &= 0
  \label{eq:Q1}\\
  H^{[i}_{\mu\, ab} H^{j]}_\nu{}^{ab} &= 0~, \label{eq:Q2}  
\end{align}
where we have used that
\begin{equation*}
  [\Gamma^{ab}\gamma_i , \Gamma^{cd}\gamma_j] =
  [\Gamma^{ab},\Gamma^{cd}]\delta_{ij} + 2 \left(\Gamma^{abcd} +
  g^{ad}g^{bc} - g^{ac}g^{bd} \right) \gamma_{ij}
\end{equation*}
and that on chiral spinors $\Gamma^{abcd} = \pm \half
\epsilon^{abcdef}\Gamma_{ef}$.

We will now analyse these equations.  Equation \eqref{eq:dH}, together
with the fact that $dH=0$, implies that $\nabla H = 0$; whence it
follows, from equation \eqref{eq:riemann}, that the Riemann curvature
tensor is covariantly constant, whence the spacetime is locally
symmetric.  Equation \eqref{eq:Q2} is identically satisfied for
anti-selfdual $H^i$.  Finally, using the anti-selfduality of $H^i$ we
can rewrite equation \eqref{eq:Q1} as
\begin{equation}
  \label{eq:Q}
  H^{[i}_{abm} H^{j]\, cd m} = 0~,
\end{equation}
where we have expressed $H^i$ in terms of a frame $\{e_a\}$.  In fact,
since the spacetime is locally symmetric, it is enough to work at a
point on whose tangent space $g$ induces a lorentzian scalar product
and $H^i$ are constant-coefficient anti-selfdual three-forms.

We claim that equation \eqref{eq:Q} implies that $\bH$ is
decomposable; that is, $H^i_{abc} = H_{abc} v^i$, where $v$ is a
unit vector and $H_{abc}$ is anti-selfdual.  If all $H^i=0$ this is
clear, so let us suppose that at least one $H^i$ is nonzero.  Without
loss of generality we can let it be $H^1$ by relabelling the frame if
necessary.  Because of anti-selfduality, we know that $\iota_0 H^1$ is
different from zero.  Being a two-form in the five-dimensional space
perpendicular to $e_0$, we can rotate in that space in such a way that
$\iota_0 H^1 = \alpha e_1\wedge e_2 + \beta e_3 \wedge e_4$, where
$\alpha \neq 0$.  Since equation \eqref{eq:Q} is homogeneous, we can
rescale $H^i$ such that $\alpha = 1$.  Finally we perform an
R-symmetry transformation on $H^i$ in such a way that $H^i_{012} = 0$
for $i\neq 1$.  In other words, by suitably changing basis we arrive
at $H^{i>1}_{012} = 0$ and $H^1_{012} =1$, $H^1_{034} = \beta$ and all
other $H^1_{0ab} = 0$.  Inserting this Ansatz into equation
\eqref{eq:Q} for $i=1$, $a=0$, $b=1$ and letting $j>1$, $c,d$ vary, we
obtain at once that $H^j=0$ for $j>1$.  In other words and
re-inserting the parameter $\alpha$, the only nonzero $H^i$ is
\begin{equation*}
  H^1 = \alpha ( e_{012} + e_{345} ) + \beta ( e_{034} + e_{125} )~,
\end{equation*}
with $\alpha \neq 0$, where we have introduced the notation $e_{012} =
e_0 \wedge e_1 \wedge e_2$, and so on.

Finally we insert this result into equation \eqref{eq:riemann} to
obtain
\begin{equation}
  \label{eq:curv20}
  R_{\mu\nu\,ab} = \tfrac18 \left( H_{\mu\, ac} H_{\nu\, db}
    - H_{\nu\, ac} H_{\mu\, db}\right) g^{cd}~,
\end{equation}
where we have let $H^1=H$.  Comparing this equation with
\eqref{eq:flat10} we see that they are indeed the same, whence there
is a one-to-one correspondence between $(2,0)$ vacua (up to
R-symmetry) and $(1,0)$ vacua.  This means, in particular, that the
$(1,0)$ vacua actually possess 16 supercharges.

More precisely, we see that any $(1,0)$ vacuum $(g,H)$ gives rise to a
$(2,0)$ vacuum $(g,\bH)$ simply by letting $\bH = H \otimes v$ with
$v$ a unit vector in $V$; that is, $H^i = H v^i$, with $v^i v^i = 1$.
Under the R-symmetry group $\Sp(2)$, $\bH$ transforms as a vector and
hence the $\Sp(2)$-transform of any vacuum is also a vacuum.  What we
have shown is that every $(2,0)$ vacuum is in the R-symmetry orbit of
a $(1,0)$ vacuum.

It thus remains to determine the $(1,0)$ vacua, a task to which we
now turn.

\section{Anti-selfdual lorentzian Lie groups}
\label{sec:lie}

In the previous section we concluded that $(1,0)$ vacua are locally
isometric to six-dimensional Lie groups admitting a bi-invariant
metric whose parallelising torsion three-form is anti-selfdual.  In
this section we determine such Lie groups up to local isometry.

A Lie group $G$ admits a bi-invariant metric (that is, a metric
invariant under both left and right multiplication) if and only if its
Lie algebra $\fg$ admits a scalar product $\left<-,-\right>$ which is
invariant (under the adjoint action):
\begin{equation*}
  \left<[X,Y],Z\right> = \left<X,[Y,Z]\right>~,
\end{equation*}
for all $X,Y,Z \in \fg$.  Relative to a basis $X_a$ for $\fg$, let
$g_{ab} = \left<X_a,X_b\right>$ and $[X_a,X_b]= f_{ab}{}^c X_c$.  Then
the invariance of the metric simply says that $f_{abc} = f_{ab}{}^d
g_{dc}$ is totally skew-symmetric.  We say that a Lie group $G$ is
\emph{lorentzian} if it has a bi-invariant lorentzian metric.
Simply-connected lorentzian Lie groups are in one-to-one
correspondence with Lie algebras admitting an invariant lorentzian
scalar product; that is $g_{ab}$ is lorentzian.  Given such a Lie
algebra there is a canonical invariant three-form $h\in\Lambda^3\fg^*$
defined by
\begin{equation*}
  h(X,Y,Z) = \left<X,[Y,Z]\right>~,
\end{equation*}
or equivalently $h_{abc} = f_{abc}$.  This form gives rise to a
bi-invariant differential three-form $H \in \Omega^3(G)$ on the Lie
group by defining it to be $h$ on left-invariant vector fields and
extending tensorially to arbitrary vector fields.  Being bi-invariant,
$H$ is both closed and co-closed. Acting on three-forms in six
dimensions (and lorentzian signature) the Hodge $\star$ operator obeys
$\star^2 = \id$, whence we can define selfdual and anti-selfdual
three-forms.  We will say that a lorentzian Lie group is
\emph{anti-selfdual} if $H$ is an anti-selfdual three-form.
Similarly, we will say that its Lie algebra is an anti-selfdual Lie
algebra.

In this section we will classify six-dimensional anti-selfdual
lorentzian Lie algebras.

\subsection{Lie algebras with an invariant metric}
\label{sec:doublext}

It is well-known that reductive Lie algebras --- that is, direct
products of semisimple and abelian Lie algebras --- admit invariant
scalar products: Cartan's criterion allows us to use the Killing forms
on the simple factors and any scalar product on an abelian Lie algebra
is trivially invariant.  Another well-known example of Lie algebras
admitting an invariant scalar product are the classical doubles.  Let
$\fh$ be \emph{any} Lie algebra and let $\fh^*$ denote the dual space
on which $\fh$ acts via the coadjoint representation.  The definition
of the coadjoint representation is such that the dual pairing $\fh
\otimes \fh^* \to \RR$ is an invariant scalar product on the
semidirect product $\fh \ltimes \fh^*$ with $\fh^*$ an abelian ideal.
The Lie algebra $\fh \ltimes \fh^*$ is called the classical double of
$\fh$ and the invariant metric has split signature $(r,r)$ where $\dim
\fh = r$.

It turns out that all Lie algebras admitting an invariant scalar
product can be obtained by a mixture of these constructions.  Let
$\fg$ be a Lie algebra with an invariant scalar product
$\left<-,-\right>_\fg$.  We will let $X_\alpha$ be a basis for $\fg$
such that $[X_\alpha,X_\beta] = f_{\alpha\beta}{}^\gamma X_\gamma$ and
such that $\left<X_\alpha,X_\beta\right> = g_{\alpha\beta}$.  Now let
$\fh$ act on $\fg$ as skew-symmetric derivations; that is, preserving
both the Lie bracket and the scalar product.  We will let $H_i$ be a
basis for $\fh$.  The action of $\fh$ on $\fg$ is given by
$f_{i\alpha}{}^\beta$, such that
\begin{equation*}
  H_i \cdot X_\alpha = f_{i\alpha}{}^\beta X_\beta~.
\end{equation*}
For future use we define $f_{i\alpha\beta} = f_{i\alpha}{}^\gamma
g_{\gamma\beta}$.  First of all, since $\fh$ acts on $\fg$ preserving
the  scalar product, we have a linear map
\begin{equation*}
  \fh \to \Lambda^2 \fg~,
\end{equation*}
with dual map
\begin{equation*}
  c: \Lambda^2 \fg \to \fh^*~,
\end{equation*}
where we have used the invariant scalar product to identity $\fg$ and
$\fg^*$ equivariantly.   Explicitly, $c(X_\alpha \wedge X_\beta) =
f_{i\alpha\beta} H^i$, where $H^i$ is the canonical dual basis of $\fh^*$.
Since $\fh$ preserves the Lie bracket in $\fg$, this map is a cocycle,
whence it defines a class $[c]\in H^2(\fg;\fh^*)$ in the second Lie
algebra cohomology of $\fg$ with coefficients in the trivial module
$\fh^*$.  Let $\fg \times_c \fh^*$ denote the corresponding central
extension.  The Lie bracket of $\fg \times_c \fh^*$ is such that
$\fh^*$ is central and if $X,Y\in\fg$, then
\begin{equation*}
  [X,Y] = [X,Y]_\fg + c(X,Y)~,
\end{equation*}
where $[-,-]_\fg$ is the original Lie bracket on $\fg$.  In terms of
the basis chosen above,
\begin{equation*}
  [X_\alpha,X_\beta] = f_{\alpha\beta}{}^\gamma X_\gamma +
  f_{i\alpha\beta} H^i~.
\end{equation*}
Now $\fh$ acts naturally on $\fg \times_c \fh^*$ preserving the Lie
bracket; the action on $\fh^*$ being given by the coadjoint
representation.  This then allows us to define the \emph{double
extension} of $\fg$ by $\fh$,
\begin{equation*}
  \fd(\fg,\fh) = \fh \ltimes (\fg \times_c \fh^*)
\end{equation*}
as a semidirect product.  Details of this construction can be found in
\cite{MedinaRevoy,FSsug}.  The remarkable fact is that $\fd(\fg,\fh)$
admits an invariant scalar product:
\begin{equation}
  \bordermatrix{& X_\beta & H_j & H^j \cr
  X_\alpha &  g_{\alpha\beta} & 0 &  0 \cr
  H_i &    0   & B_{ij} & \delta_i^j \cr
  H^i &    0   & \delta_j^i &  0 \cr}
\end{equation}
where $B$ is \emph{any} invariant symmetric bilinear form on $\fh$.

We say that a Lie algebra with an invariant scalar product is
\emph{indecomposable} if it cannot be written as the direct product of
two orthogonal ideals.  A theorem of Medina and Revoy
\cite{MedinaRevoy} (see also \cite{FSalgebra} for a refinement) says
that an indecomposable (finite-dimensional) Lie algebra with an
invariant scalar product is one of the following:
\begin{enumerate}
\item one-dimensional,
\item simple, or
\item a double extension $\fd(\fg,\fh)$ where $\fh$ is either simple
  or one-dimensional and $\fg$ is a (possibly trivial) Lie algebra
  with an invariant scalar product.
\end{enumerate}
Any (finite-dimensional) Lie algebra with an invariant scalar product
is then a direct sum of indecomposables.

\subsection{Anti-selfdual lorentzian Lie algebras}

Notice that if the scalar product on $\fg$ has signature $(p,q)$ and
if $\dim \fh = r$, then the scalar product on $\fd(\fg,\fh)$ has
signature $(p+r,q+r)$.  Therefore indecomposable lorentzian Lie
algebras are either reductive or double extensions $\fd(\fg,\fh)$
where $\fg$ has a positive-definite invariant scalar product and $\fh$
is one-dimensional.  In the reductive case, indecomposability means
that it has to be simple, whereas in the latter case, since the scalar
product on $\fg$ is positive-definite, $\fg$ must be reductive.  A
result of \cite{FSsug} (see also \cite{FSalgebra}) then says that any
semisimple factor in $\fg$ splits off resulting in a decomposable Lie
algebra.  Thus if the double extension is to be indecomposable, then
$\fg$ must be abelian.  In summary, an indecomposable lorentzian Lie
algebra is either simple or a double extension of an abelian Lie
algebra by a one-dimensional Lie algebra and hence solvable (see,
e.g., \cite{MedinaRevoy}).

These considerations make possible the following enumeration of
six-dimensional lorentzian Lie algebras:
\begin{enumerate}
\item $\EE^{1,5}$
\item $\EE^{1,2} \oplus \fso(3)$
\item $\EE^3 \oplus \fso(1,2)$
\item $\fso(1,2) \oplus \fso(3)$
\item $\fd(\EE^4,\RR)$
\end{enumerate}
where the last case actually corresponds to a family of Lie algebras,
depending on the action of $\RR$ on $\RR^4$, which is given by a
homomorphism $\RR \to \fso(4)$.

Imposing the condition of anti-selfduality trivially discards cases
(2) and (3) above.  Case (1) is the abelian Lie algebra with Minkowski
metric.  We now investigate in more detail the remaining two cases,
starting with case (5) which requires more attention.

\subsubsection{A six-dimensional Nappi--Witten vacuum}

Let $e_i$, $i=1,2,3,4$, be an orthonormal basis for $\EE^4$, and let
$e_- \in \RR$ and $e_+\in\RR^*$, so that together they span
$\fd(\EE^4,\RR)$.  The action of $\RR$ on $\EE^4$ defines a map $\rho:
\RR \to \Lambda^2 \EE^4$, which can be brought to the form $\rho(e_-) =
\alpha e_1 \wedge e_2 + \beta e_3 \wedge e_4$ via an orthogonal change
of basis in $\EE^4$ which moreover preserves the orientation.  The Lie
brackets of $\fd(\EE^4,\RR)$ are given by
\begin{equation*}
  \begin{aligned}[m]
    [e_-,e_1] &= \alpha e_2\\
    [e_-,e_2] &= -\alpha e_1\\
    [e_1,e_2] &= \alpha e_+
  \end{aligned}\qquad\qquad
  \begin{aligned}[m]
    [e_-,e_3] &= \beta e_4\\
    [e_-,e_4] &= -\beta e_3\\
    [e_3,e_4] &= \beta e_+
  \end{aligned}~,
\end{equation*}
and the scalar product is given (up to scale) by
\begin{equation*}
  \left<e_-,e_-\right> = b \qquad \left<e_+,e_-\right> = 1 \qquad
  \left<e_i,e_j\right> = \delta_{ij}~.
\end{equation*}
The first thing we notice is that we can set $b=0$ without loss of
generality by the automorphism fixing all $e_i,e_+$ and mapping $e_-
\mapsto e_- - \half b e_+$.  We will assume that this has been done
and that $\left<e_-,e_-\right>=0$.  A straightforward calculation
shows that the three-form $f_{abc}$ is anti-selfdual if and only if
$\beta = \alpha$.  Let us put $\beta=\alpha$ from now on.  We must
distinguish between two cases: if $\alpha = 0$, then the resulting
algebra is abelian and is precisely $\EE^{1,5}$.  On the other hand if
$\alpha\neq 0$, then rescaling $e_\pm \mapsto \alpha^{\pm 1} e_\pm$ we
can effectively set $\alpha = 1$ without changing the scalar product.
Finally we notice that a constant rescaling of the scalar product can
be undone by an automorphism of the algebra.  As a result we have two
cases: $\EE^{1,5}$ (obtained from $\alpha =0$) and the algebra
\begin{equation*}
  \begin{aligned}[m]
    [e_-,e_1] &= e_2\\
    [e_-,e_2] &= - e_1\\
    [e_1,e_2] &= e_+
  \end{aligned}\qquad\qquad
  \begin{aligned}[m]
    [e_-,e_3] &= e_4\\
    [e_-,e_4] &= - e_3\\
    [e_3,e_4] &= e_+
  \end{aligned}~,
\end{equation*}
with scalar product given by
\begin{equation*}
  \left<e_+,e_-\right> = 1 \qquad\text{and}\qquad
  \left<e_i,e_j\right> = \delta_{ij}~.
\end{equation*}
There is a unique simply-connected Lie group with the above Lie
algebra which inherits a bi-invariant lorentzian metric.  This Lie
group is a six-dimensional analogue of the Nappi--Witten group
\cite{NW}, which is based on the double extension $\fd(\RR^2,\RR)$
\cite{FSsug}.  We will denote it $\NW_6$ as in \cite{FSPL}, where one
can find a derivation of the metric on this six-dimensional group.  It
can be seen to be a symmetric plane wave (Hpp-wave in the terminology
of \cite{FOPflux}) corresponding to a symmetric space of the type
discovered by Cahen and Wallach \cite{CahenWallach}.  The supergravity
solution was discovered by Meessen \cite{Meessen} who called it KG6 by
analogy with the maximally supersymmetric plane wave of
eleven-dimensional supergravity discovered by Kowalski-Glikman
\cite{KG} (see also \cite{FOPflux}).

The metric is easy to write down once we choose a parametrisation for
the group.  The calculation is routine (see, for example, \cite{FSPL})
and the result is
\begin{equation}
  \label{eq:CWmetric}
  g = 2 dx^+ dx^- - \tfrac14 \sum_i (x^i)^2 (dx^-)^2 + \sum_i
  (dx^i)^2~.
\end{equation}
In these coordinates the three-form $H$ is given by
\begin{equation*}
  H = \tfrac23 dx^- \wedge (dx^1 \wedge dx^2 + dx^3 + dx^4)~.
\end{equation*}

The metric \eqref{eq:CWmetric} corresponds to a lorentzian symmetric
space of the type introduced by Cahen and Wallach \cite{CahenWallach}
and discussed more recently in the context of plane wave solutions of
supergravity theories in \cite{FOPflux}.  However in the present
context it appears as a bi-invariant metric on a solvable
group.\footnote{Every Lie group $G$ with a bi-invariant metric is
  isometric to the symmetric space $(G\times G)/\Delta G$, where
  $\Delta G$ is the diagonal $G$ subgroup of $G\times G$.  However
  this is not the same symmetric space in the description of Cahen and
  Wallach.}  This is not an isolated incident.  In fact, it is not
hard to characterise those Cahen--Wallach plane wave metrics which are
isometric to a bi-invariant metric on a solvable Lie group.  Recall
that an indecomposable Cahen--Wallach metric takes the form
\begin{equation*}
  g = 2 dx^+ dx^- + \sum_{i,j=1}^n A_{ij} x^i x^j (dx^-)^2 +
  \sum_{i=1}^n (dx^i)^2~,
\end{equation*}
where the symmetric matrix $A$ is nondegenerate.  It is proven in
\cite{FSPL} that $g$ is isometric to a bi-invariant metric on a
solvable Lie group if and only if $A$ is negative-definite and all its
eigenvalues have even multiplicity.  This means that $A$ admits a
decomposition $A = J^2$, where $J$ is skew-symmetric and
nondegenerate, and the Lie algebra is a double extension of $\EE^n$ by
$\RR$, where the generator of $\RR$ acts on $\EE^n$ by $J$.  Notice
that, as observed in \cite{FSPL}, this implies that the IIB maximally
supersymmetric wave \cite{NewIIB} is isometric to a solvable Lie group
with a bi-invariant metric, whereas the maximally supersymmetric
M-wave \cite{KG} is not.

\subsubsection{The Freund--Rubin vacuum}

Finally we discuss case (4), with Lie algebra $\fso(1,2) \oplus
\fso(3)$.  Let $e_0,e_1,e_2$ be a pseudo-orthonormal basis for
$\fso(1,2)$.  The Lie brackets are given by
\begin{equation*}
  [e_0,e_1] = -e_2\qquad   [e_0,e_2] = e_1\qquad   [e_1,e_2] =
  e_0~.
\end{equation*}
Similarly let $e_3,e_4,e_5$ denote an orthonormal basis for
$\fso(3)$, with Lie brackets
\begin{equation*}
  [e_5,e_3] = -e_4\qquad   [e_5,e_4] = e_3\qquad   [e_3,e_4] =
  -e_5~.
\end{equation*}
The most general invariant lorentzian scalar product on
$\fso(1,2)\oplus\fso(3)$ is labelled by two positive numbers $\alpha$
and $\beta$ and is given by
\begin{equation*}
  \bordermatrix{& e_0 & e_1 & e_2 & e_3 & e_4 & e_5 \cr
  e_0 &  -\alpha & 0  &  0  & 0 & 0 & 0 \cr
  e_1 & 0 & \alpha & 0  & 0 & 0 & 0 \cr
  e_2 & 0 & 0 & \alpha & 0 & 0 & 0 \cr
  e_3 & 0 & 0 & 0 & \beta & 0 & 0 \cr
  e_4 & 0 & 0 & 0 & 0 & \beta &  0 \cr
  e_5 & 0 & 0 & 0 & 0 & 0 & \beta \cr}~.  
\end{equation*}
Anti-selfduality of the canonical three-form implies that $\beta =
\alpha$.  There is a unique simply-connected Lie group with Lie
algebra $\fso(1,2) \oplus \fso(3)$, namely $\widetilde{\SL(2,\RR)}
\times \SU(2)$, where $\widetilde{\SL(2,\RR)}$ denotes the universal
covering group of $\SL(2,\RR)$.  This group inherits a one-parameter
family of bi-invariant metrics.  This solution is none other than the
standard Freund--Rubin solution $\AdS_3 \times S^3$, with equal radii
of curvature, where strictly speaking we should take the universal
covering space of $\AdS_3$.

In summary, the following are the possible vacua of $(1,0)$ and (up to
R-symmetry) $(2,0)$ supergravity.  First of all we have a
one-parameter family of Freund-Rubin vacua locally isometric to
$\AdS_3 \times S^3$, with equal radii of curvature.  The anti-selfdual
three-form $H$ is then proportional to the difference of the volume
forms of the two spaces.

Then we have a six-dimensional analogue $\NW_6$ of the Nappi--Witten
group, locally isometric to a Cahen--Wallach symmetric space.  Finally
there is the flat vacuum $\RR^{1,5}$.  These vacua are related by
Penrose limits which can be interpreted as group contractions.  The
details appear in \cite{FSPL}.

\section{Five-dimensional vacua and Kaluza--Klein reduction}
\label{sec:vacua5}

In this section we will examine the dimensional reductions of the
six-dimensional vacua found above.  Dimensional reduction usually
breaks some supersymmetry: in the ten- and eleven-dimensional
supergravity theories, only the flat vacuum remains maximally
supersymmetric after dimensional reduction and only by a translation.
However for the six-dimensional vacua the situation is different.
Indeed, in \cite{LMO8} it was shown that the hitherto known
supergravity vacua with eight supercharges in six, five and four
dimensions are related by dimensional reduction and oxidation.  As we
will see presently, this perhaps surprising phenomenon follows from
the fact that the six-dimensional vacua are parallelised Lie groups.
Our results will also give an \emph{a priori} explanation to the
empirical fact that these vacua are homogeneous \cite{ALO}.

\subsection{Geometric preliminaries}

We start this section with a technical result which underlies the rest
of the section.  Let $D$ be a metric connection with torsion $T$.  We
observe that if a vector field $\xi$ is $D$-parallel then it is
Killing.  Indeed, $D\xi = 0$ implies that $\nabla_\mu\xi_\nu = \half
\xi^\rho H_{\mu\nu\rho}$, where $H$ is the associated torsion
three-form.  Therefore we see that $\nabla_\mu \xi_\nu = - \nabla_\nu
\xi_\mu$, whence $\xi$ is Killing.  Now let $\psi$ be a Killing
spinor; that is, $D\psi = 0$. Then the Lie derivative of $\psi$ along
$\xi$ is well-defined (see, for example, \cite{JMFKilling}).
Furthermore, it vanishes identically.  Indeed, by definition,
\begin{equation*}
  \begin{aligned}[m]
    \eL_\xi \psi &= \nabla_\xi \psi + \tfrac14 \nabla_{[\mu} \xi_{\nu]}
    \Gamma^{\mu\nu} \psi\\
    &= \tfrac18 \left( \nabla_\mu \xi_\nu - \nabla_\nu \xi_\mu -
      \xi^\rho H_{\mu\nu\rho} \right) \Gamma^{\mu\nu} \psi~,
  \end{aligned}
\end{equation*}
where we have used that $D_\xi \psi = 0$.  Since $\xi$ is Killing,
$\nabla_\nu \xi_\mu = - \nabla_\mu \xi_\nu$, whence
\begin{equation*}
  \eL_\xi \psi = \tfrac14 \left( \nabla_\mu \xi_\nu - \half \xi^\rho
    H_{\mu\nu\rho} \right) \Gamma^{\mu\nu} \psi~,
\end{equation*}
which vanishes when $D\xi=0$.  Moreover it follows from the above
calculation that if $\eL_\xi \psi = 0$ for \emph{all} Killing spinors
then $\nabla_\mu \xi_\nu = \half \xi^\rho H_{\mu\nu\rho}$, so that
$D\xi = 0$.

For a parallelised Lie group $G$, the $D$-parallel vectors are either
the left- or right-invariant vector fields, depending on the choice of
parallelising connection.  For definiteness, we will choose the
connection whose parallel sections are the left-invariant vector
fields.  Left-invariant vector fields generate right translations and
are in one-to-one correspondence with elements of the Lie algebra
$\fg$.  Therefore every left-invariant vector field $\xi$ determines a
one-parameter subgroup $K$, say, of $G$ and the orbits of such a
vector field in $G$ are the right $K$-cosets.  The dimensional
reduction along this vector field is smooth and diffeomorphic to the
space of cosets $G/K$.  We will be interested in subgroups $K$ such
that $G/K$ is a five-dimensional lorentzian spacetime, which requires
that the right $K$-cosets are spacelike.  In other words, we require
that the Killing vector $\xi$ be spacelike.  Bi-invariance of the
metric guarantees that this is the case provided that the Lie algebra
element $\xi(e)\in \fg$ is spacelike relative to the ad-invariant
inner product.  Further notice that a constant rescaling of $\xi$ does
not change its causal property nor the subgroup $K$ it generates: it
is simply reparameterised.  Therefore, in order to classify all
possible reductions (and hence all possible five-dimensional vacua
with 8 supercharges) we need to classify all spacelike elements of
$\fg$ up to scale.  Moreover elements of $\fg$ which are related by
isometric automorphisms (e.g., which are in the same adjoint orbit of
$G$) give rise to isometric quotients.  Thus, to summarise, we want to
classify spacelike elements of $\fg$ up to scale and up to
automorphisms.  For a more detailed explanation of this reasoning, the
reader is referred to the papers \cite{FigSimFlat, FigSimBranes},
which also contain the description of the geometric set-up for
Kaluza--Klein reduction which we now briefly review.

Let $X\in\fg$ be spacelike and let $\xi_X$ denote the corresponding
left-invariant vector field.  By rescaling $X$ if necessary we can
always take $X$ (and hence $\xi_X$) to have unit norm.  Let $K$ be the
one-parameter subgroup of $G$ generated by $X$.  The natural map
$\pi:G \to G/K$, taking a group element to the right $K$-coset it
belongs to, is actually a principal fibration with group $K$.  The
tangent space $T_g G$ has a canonical subspace $\eV_g = \ker \pi_*$
corresponding to the span of $\xi_X(g)$.  Because $\xi_X(g)$ has
non-vanishing norm, there is a decomposition
\begin{equation*}
  T_g G = \eV_g \oplus \eH_g
\end{equation*}
where $\eH_g = \eV_g^\perp$ consists of all those tangent vectors at
$g$ which are perpendicular to $\xi_X(g)$.  Because the metric in $G$
is bi-invariant, we can identify $\eH_g$ with the left-translate (by
$(L_g)_*$) of those vectors in $\fg$ which are perpendicular to $X$.
The distribution $\eH$ is a connection in the sense of Ehresmann and
has an associated connection one-form $\alpha$ in $G$, defined
by $\ker \alpha = \eH$ and normalised to $\alpha(\xi_X) = 1$.
Explicitly, $\alpha = \xi^\flat_X/\|\xi_X\|^2$, where $\xi^\flat_X$
is the one-form dual to $\xi_X$.  With our choice of normalisation for
the Killing vector, $\|\xi_X\| = \|X\| = 1$, whence $\alpha =
\xi^\flat_X$.  We can give an even more explicit expression for
$\alpha$ in terms of the left-invariant Maurer--Cartan one-form
$\theta$.  Indeed, $\alpha = \left<X,\theta\right>$, where
$\left<-,-\right>$ is the metric in $\fg$.  To see this, notice that
if $\zeta$ is a vector field in $G$, then $\theta(\zeta)(g) =
(L_{g^{-1}})_* \zeta(g) \in T_e G = \fg$.  Therefore,
\begin{equation*}
  \begin{aligned}[m]
    \alpha(\zeta)(g) &= \left<X,\theta(\zeta)(g)\right>\\
                     &= \left<X,(L_{g^{-1}})_* \zeta(g)\right>\\
                     &= \left<(L_g)_* X, \zeta(g)\right>_g\\
                     &= \left<\xi_X(g), \zeta(g)\right>_g\\
                     &= \xi_X^\flat(\zeta)(g)~,
  \end{aligned}
\end{equation*}
where in the third line we use the left-invariance of the metric, in
the fourth we have used the left-invariance of $\xi_X$, and we have
introduced the notation $\left<-,-\right>_g$ to be the metric at $g\in
G$ with $\left<-,-\right>_e := \left<-,-\right>$ the one in the Lie
algebra.

The reduction of the six-dimensional metric to five dimensions gives
rise to several geometric structures (see, for example,
\cite{FigSimFlat,FigSimBranes}): a metric $h$, a dilaton $\phi$ and a
2-form field strength $F$.  The metric $h$ is the induced metric on the
horizontal distribution $\eH$, the dilaton $\phi$ is a logarithmic
measure of the fibre metric $\|\xi_X\|$ which in our case is constant,
and $F = d\alpha$.  We can give an explicit formula for $F$ using the
Maurer--Cartan structure equations.  Indeed,
\begin{equation}
  \label{eq:F}
  F = d\alpha = \left<X,d\theta\right> = -\half
  \left<X,[\theta,\theta]\right>~.
\end{equation}
In terms of this data, the metric on the $G$ is given by the usual
Kaluza--Klein ansatz
\begin{equation*}
  ds^2 = h + \alpha^2~,
\end{equation*}
where we have set the dilaton to zero in agreement with the choice of
normalisation for $\xi_X$.  More explicitly the metric on the
five-dimensional quotient is given by
\begin{equation*}
  h = \left<\theta,\theta\right> - \left<X,\theta\right>^2~.
\end{equation*}
In other words, if $e_\mu$ is a basis for the perpendicular complement
of $X$ in $\fg$, with $\left<e_\mu,e_\nu\right> = \eta_{\mu\nu}$ then the
metric on the five-dimensional quotient is
\begin{equation*}
  h = \eta^{\mu\nu} \left<e_\mu,\theta\right>
  \left<e_\nu,\theta\right>~.
\end{equation*}

To reduce the anti-selfdual three-form $H$ we first decompose it as
\begin{equation*}
  H = G_3 + \alpha \wedge G_2~,
\end{equation*}
where $G_2 = \iota_{\xi_X} H$ and $G_3$ are horizontal; that is,
$\iota_{\xi_X} G_3 = \iota_{\xi_X} G_2 = 0$.  Because the Killing
vector $\xi_X$ leaves $H$ invariant, it follows that also $G_2$ and
$G_3$ are invariant; that is, $\eL_{\xi_X}G_2 = \eL_{\xi_X}G_3 = 0$.
In other words, $G_2$ and $G_3$ are \emph{basic}; that is, they are
pullbacks of forms in the base $G/K$, which we will denote by the same
letters.  Because $dH=0$ it follows that $dG_2=0$ and that $dG_3 + F
\wedge G_2=0$ where $F = d\alpha$ was defined above.  Finally because
$H$ is anti-selfdual, it follows that $G_3$ and $G_2$ are related by
Hodge duality in five dimensions: $G_3 = \star_5 G_2$.  In other
words, we have that
\begin{equation*}
  H = \star_5 G_2 + \alpha \wedge G_2~,
\end{equation*}
where $dG_2 = 0$ and $d\star_5 G_2 = -F \wedge G_2$.

In fact, in this case we have $F=G_2$, whence we are dealing with
reductions which truncate consistently to the minimal five-dimensional
supergravity theory.  Recall that the dimensional reduction of $(1,0)$
supergravity to five dimensions is $N{=}2$ supergravity coupled to a
vector multiplet.  The minimal $N{=}2$ supergravity is obtained by
setting the fields in the vector multiplet to zero.  These fields are
the ``dilaton'' whose exponential is the fibre metric and a two-form
field strength which is the difference $F-G_2$.  To show that $F -
G_2$ vanishes in these reductions, we simply use that $H = -\tfrac16
\left<\theta,[\theta,\theta]\right>$ and compute
\begin{equation*}
   G_2 = \iota_{\xi_X} H = -\half \left<X,[\theta,\theta]\right>~,
\end{equation*}
which agrees with the expression for $F$ derived in \eqref{eq:F}.
This truncation is consistent with supersymmetry since by construction
the supersymmetry variations of the fields in the vector multiplet
also vanish.

In summary, for the reductions under consideration, we obtain a vacuum
of the minimal $N{=}2$ supergravity with bosonic fields $(h,F)$ given
by the reduction of $(g,H)$ where $F=d\alpha$, $h = g - \alpha^2$ and
$H = \star_5 F + \alpha \wedge F$.

\subsection{Possible Kaluza--Klein reductions}

We now classify the possible Kaluza--Klein reductions to five
dimensions, by classifying the spacelike one-parameter subgroups of
the Lie groups in question.  As outlined above this is achieved by
classifying the normal forms of elements of the Lie algebra $\fg$
under rescalings and metric-preserving automorphisms.

\subsubsection{Spacelike subgroups of $\RR^{1,5}$}

The Lie algebra is abelian and hence all automorphisms are outer.  The
metric-preserving automorphism group is $\Ort(1,5)$ acting in the
obvious way.  Up to $\Ort(1,5)$ any spacelike element in $\RR^{1,5}$
can be rotated so that it generates translation along the fifth
spatial coordinate $x^5$.  The resulting quotient is $\RR^{1,4}$ with
vanishing fluxes and constant dilaton.

\subsubsection{Spacelike subgroups of $\widetilde{\SL(2,\RR)} \times
\SU(2)$}
\label{sec:adsxsreds}

The Lie algebra is now $\fso(1,2) \oplus \fso(2,3)$.  Every nonzero
element of $\fso(3)$ is conjugate under $\SO(3)$ to any other nonzero
element of the same norm, whereas nonzero elements of $\fso(1,2)$ come
in three flavours under $\SO(1,2)$: elliptic, parabolic or hyperbolic,
depending on whether it has positive, zero or negative norm,
respectively.  Let $\sigma$, $\nu$ and $\tau$ denote respectively a
spacelike, null or timelike element in $\fso(1,2)$.  Let us normalise
$\sigma$ and $\tau$ such that $\|\sigma\|^2 =1$ and $\|\tau\|^2 = -1$.
Let $\kappa$ denote any unit-norm element in $\fso(3)$.  Then we have
three types of spacelike elements in $\fso(1,2) \oplus \fso(3)$:
\begin{enumerate}
\item $\xi = a \sigma + b \kappa$, where $a$,$b$ are not both zero;
\item $\xi = \nu + b \kappa$, where $b$ is not zero; and
\item $\xi = a \tau + b \kappa$, where $b^2 > a^2$.
\end{enumerate}
Rescaling and using the fact that in case (2) we can renormalise the
coefficient of $\nu$ to unit via an $\SO(1,2)$ transformation, we
obtain
\begin{enumerate}
\item $\xi = \frac{a \sigma + \kappa}{\sqrt{1+a^2}}$, where $a$ is
  arbitrary;
\item[(1')] $\xi = \sigma$;
\item $\xi = \nu + \kappa$; and
\item $\xi = \frac{a \tau + \kappa}{\sqrt{1-a^2}}$, where $-1<a<1$.
\end{enumerate}
Notice that we have split the previous case (1) into two and we have
normalised the Killing vector so that it has unit norm.

\subsubsection{Spacelike subgroups of $\NW_6$}

The metric-preserving automorphisms of the Lie algebra $\fn$ of
$\NW_6$ have been determined in \cite{FSPL} and they define a group
isomorphic to $\RR^4 \rtimes (\U(2) \rtimes \ZZ_2)$, where $\U(2)
\rtimes \ZZ_2$ is the principal extension of $\U(2)$ by the outer
automorphism consisting of complex conjugation, and acts on $\RR^4
\cong \CC^2$ as follows:
\begin{equation*}
  (U,1) \cdot \bz = U\bz \qquad\text{and}\qquad (U,-1) \cdot \bz = - U
  \bar \bz~,
\end{equation*}
where $U\in \U(2)$ and $\bz \in \CC^2$.  Let us consider a spacelike
vector in the Lie algebra $\fn$:
\begin{equation*}
  \bv + v^- e_- + v^+ e_+ \qquad\text{with}\quad \|\bv\|^2 +
  2 v^+ v^- > 0~,
\end{equation*}
where $\bv = v^i e_i$ but we think of it as a vector in $\CC^2$ by
taking explicit complex linear combinations $e_1 + i e_2$ and $e_3 + i
e_4$.  The action of $(\bz, U, 1) \in \CC^2 \rtimes (\U(2) \rtimes
\ZZ_2)$ on such a vector is given by
\begin{equation*}
  (\bz, U,1) \cdot 
  \begin{pmatrix}
    \bv\\ v^-\\ v^+
  \end{pmatrix}
  = 
  \begin{pmatrix}
    U \bv - v^- \bz\\
    v^-\\
    \bar\bz^t U \bv - \half \|\bz\|^2 v^- + v^+
   \end{pmatrix}~.
\end{equation*}
We claim that we can always put $v^+$ and all components of $\bv$
except for $v^1$, say, to zero via such an automorphism.  There are
three cases to consider:
\begin{enumerate}
\item $\bv\neq0$ and $v^-\neq 0$: In this case we act with
  $(\bzero,U,1)$ to set all components of $\bv$ to zero except for
  $v^1$, say.  Then we act with $(\bz,\1,1)$ to set $v^+ =0$ with an
  appropriate $\bz$.  Such a $\bz$ exists \emph{precisely} because the
  vector is spacelike, and moreover it can be chosen so that $\bv$
  remains with all components zero except for $v^1$.
\item $v^-=0$ and $\bv \neq 0$: Here we act with $(\bz,\1,1)$, for an
  appropriate $\bz$, to set $v^+ = 0$ and such that the resulting
  $\bv$ has all components zero except for $v^1$.  This is clearly
  possible.
\item $\bv=0$ and $v^-\neq 0$: Again we act with $(\bz,\1,1)$, for an
  appropriate $\bz$, to set $v^+ = 0$ and such that the resulting
  $\bv$ has all components zero except for $v^1$, which is again
  clearly possible.
\end{enumerate}
We summarise this by observing that after a possible rescaling we can
always bring a spacelike vector in $\fn$ to the form $e_1 + a e_-$,
where $a$ is some arbitrary real parameter.

\subsection{Five-dimensional vacua}

Each of the Kaluza--Klein reductions of the previous section gives
rise to a vacuum solution of the minimal $N{=}2$ supergravity theory.
In \cite[Section 5]{GGHPR} the authors studied the maximally
supersymmetric backgrounds of this supergravity theory.  Their results
include a list of vacua: flat space, a hitherto unknown Gödel-like
universe, Meessen's five-dimensional wave, and a one-parameter family
of vacua \cite{GMT} interpolating between $\AdS_2 \times S^3$ and
$\AdS_3 \times S^2$, which can be understood as the near-horizon
geometries of the supersymmetric rotating black holes of \cite{KRW}.
In addition to this list there are three other solutions which were
not yet identified or indeed shown to be maximally supersymmetric.
A detailed comparison between our results is hindered by the
intrinsic difficulty in comparing metrics which are written in terms
of local coordinates.  Nevertheless one can hazard a sort of
correspondence between our results and those of \cite[Section
5]{GGHPR}.

First of all, flat space is of course the unique maximally
supersymmetric reduction of flat space.  The near-horizon geometries
of the supersymmetric rotating black holes coincide with the
reductions (1) and (1') of the $\AdS_3 \times S^3$ vacuum.  Reductions
(2) and (3) in Section~\ref{sec:adsxsreds} probably correspond to
``near-horizon'' limits of the supersymmetric over-rotating black
holes; that is, in the regimes where the angular momentum exceeds the
physical bound.  Such solutions have closed timelike curves and this
is consistent with observed phenomena in the similar reductions in
\cite{FigSimBranes} where the Killing vector generating the reduction,
though being spacelike, is a linear combination of a spacelike and a
causal Killing vector.\footnote{Joan Simón, private communication.}  In
particular, this would mean that our reduction (2) agrees with the
solution given by equation (5.102) in \cite{GGHPR}.  Finally both the
Gödel and plane wave backgrounds arise as reductions of the
six-dimensional wave\footnote{Originally this is an unpublished result
  of the authors of \cite{LMO8}.}; and the parameter in the reductions
of Meessen's plane wave is probably related to the parameter in the
Gödel solution.

Although we have proven that all our reductions are maximally
supersymmetric, it remains to identify them.  This is work in progress
and will be presented elsewhere.

\section{Conclusions and summary of results}
\label{sec:conc}

In this paper we have classified (up to local isometry) the maximally
supersymmetric solutions (vacua) of (1,0) and (2,0) supergravities in
six dimensions and of the minimal $N{=}2$ supergravity in five
dimensions.  The (1,0) vacua are in one-to-one correspondence with
six-dimensional Lie groups with a bi-invariant lorentzian metric and
with anti-selfdual parallelising torsion.  These are easily
classified using the known results on lorentzian Lie groups, and we
have seen that all vacua are locally isometric to the known vacua:
flat space, $\AdS_3 \times S^3$ or Meessen's symmetric plane wave.
Moreover we have also proven that the (2,0) vacua are---up to the
action of the R-symmetry group $\Sp(2)$---in one-to-one correspondence
with the (1,0) vacua.  Finally we have shown that the $N{=}2$
five-dimensional vacua are spaces of (right) cosets of the above Lie
groups by one-parameter subgroups generated by left-invariant vector
fields.  We have therefore classified the possible reductions by
classifying the inequivalent such one-parameter subgroups.

\section*{Acknowledgments}

It is a pleasure to thank Tomás Ortín and Joan Simón for useful
conversations.  This work was started while JF was visiting CAMS and
it is his pleasure to thank his two co-authors for the invitation, and
the Royal Society and CAMS for the support which made the visit
possible.  JF would also like to thank the Rutgers NHETC and the
School of Natural Science of the IAS for hospitality and support
during the final stages of this work and for the opportunity to give
seminars on these results.  The research of JF is partially funded by
the EPSRC grant GR/R62694/01.  JF is a member of EDGE, Research
Training Network HPRN-CT-2000-00101, supported by The European Human
Potential Programme.

\bibliographystyle{utphys}
\bibliography{AdS,AdS3,ESYM,Sugra,Geometry,CaliGeo}

\end{document}